\documentclass[a4paper,11pt]{article}
\pdfoutput=1 % if your are submitting a pdflatex (i.e. if you have
\usepackage{jcappub} % for details on the use of the package, please
\usepackage[T1]{fontenc} % if needed

\title{Comment on `Measuring the Hubble Constant Using Strongly Lensed Gravitational Wave Signals'}

\author[a]{J.S.C. Poon,}
\author[a]{O.A. Hannuksela}

\affiliation[a]{Department of Physics, The Chinese University of Hong Kong (CUHK),\\Shatin, New Territories, Hong Kong}

\emailAdd{jason.poon@link.cuhk.edu.hk}
\emailAdd{hannuksela@phy.cuhk.edu.hk}

\abstract{In Huang \textit{et al} (JCAP08(2023)003), the work claimed that the Hubble constant can be constrained from lensed gravitational waves, despite the absence of information on GW source redshift or Einstein radius (angular size of the lens), as long as lens redshift is provided independently. To demonstrate, the authors set up a singular isothermal sphere lens model and applied Fisher information matrix and Bayesian parameter estimation on injected lensed GWs. However, their conclusion contradicts our previous work on degeneracies of GW lens system parameters. In this comment, we illustrate that Hubble constant cannot be measured with lens redshift alone, in the absence of source redshift or (Einstein radius).}

\notoc
\counterwithout{equation}{section}
\begin{document}
\maketitle

In \cite{Huang_2023}, it is stated that one can obtain meaningful constraints on the Hubble constant $H_0$ with the information of the lens redshift $z_L$, without the source redshift $z_S$ or the Einstein radius $\theta_E$ (angular size of lens) being known. 
In particular, the work shows, using Fisher information matrix (FIM) and Bayesian parameter estimation (in Fig. 4), that the lensed gravitational waves (GWs) can pinpoint the Hubble constant $H_0$ to the level of 1$\%$.
Here we argue, that it is impossible to measure the Hubble constant without constraints on source redshift (or Einstein radius). 

In particular, following Section II in \cite{Huang_2023}: the GW is lensed by a point mass lens (with mass $M_L$ and redshift $z_L$) and produced two images. 
Lens redshift $z_L$ is measured independently, leaving the source position $\eta_L$, lens mass $M_L$, source redshift $z_S$ and Hubble constant $H_0$ being the unknown parameters.
The magnification of the two images ($\mu_\pm$) and the time delay between the two images $\Delta t_d$ are given by \cite{Huang_2023}:
\begin{equation}
    \mu_{\pm}=\frac{1}{2}\pm\frac{y^2+2}{2y\sqrt{y^2+4}},
    \label{mu point mass}
\end{equation}
\begin{equation}
    \Delta t_d = 4M_L(1+z_L)\left(\frac{y\sqrt{y^2+4}}{2}+\ln{\frac{\sqrt{y^2+4}+y}{\sqrt{y^2+4}-y}}\right),
    \label{td point mass}
\end{equation}
where $y=\frac{\eta_L}{2\sqrt{M_L}}\sqrt{\frac{D_L}{D_S D_{LS}}}$ is the dimensionless source position (with $\eta_L$ being the physical source position in distance unit, $D_L, D_S, D_{LS}$ being the angular diameter distance to the lens, source and between lens and source respectively).

From the observed lensed GWs, we can measure the time delay $\Delta t_d$ and the apparent/effective luminosity distance $d_{\text{Lum, effective}}={d_\text{Lum}}/{\sqrt{|\mu |}}$ directly (where $d_\text{Lum}$ is the original/unlensed luminosity distance) \cite{poon2024galaxylensreconstructionbased}. 
Effectively, the claim in \cite{Huang_2023} is that we can measure the Hubble constant $H_0$ and source redshift $z_S$ simultaneously, with only lens redshift $z_L$ provided independently (for example, from spectroscopic measurement).

Here we show that the Hubble constant and source redshift remain degenerate, with the point mass lens example presented in \cite{Huang_2023}, when only lens redshift is obtained independently. 
In particular, consider a perfect measurement of the effective luminosity distances and the lensing time delay. 
In that case, we can obtain the relative magnification of the signals using the measured effective luminosity distances, that is $|\mu_+/\mu_-|=(d_{\text{Lum, effective},-}/d_{\text{Lum, effective},+})^2$, and use it to solve the dimensionless source position $y$ (by Eq. \ref{mu point mass}). 
We can then calculate the absolute magnification of the positive and negative parity images $\mu_+$ and $\mu_-$, and obtain the original luminosity distance by $d_\text{Lum}=d_{ \text{Lum,effective}}\sqrt{|\mu|}$. 
In addition, with the independently measured lens redshift $z_L$, we can also obtain the lens mass $M_L$ from Eq. \ref{td point mass} \footnote{We note that having measurement on the lens mass does not imply having measurement on the Einstein radius, as Einstein radius generally depends on Hubble constant and the lens and source redshift as well.}.
Therefore, we have two remaining equations to solve for three unknowns ($\eta_L, z_S, H_0$):
\begin{equation}
\begin{cases}
y=\frac{\eta_L}{2\sqrt{M_L}}\left[H_0\frac{(1+z_S)^2}{1+z_L}\frac{\int_0^{z_L}{\frac{dz'}{E(z')}}}{\int_0^{z_S}{\frac{dz'}{E(z')}}\int_{z_L}^{z_S}{\frac{dz'}{E(z')}}}\right]^{1/2}
\\
d_\text{Lum} = \frac{1+z_S}{H_0}\int_0^{z_S}{\frac{dz'}{E(z')}}
\end{cases}.
\end{equation}
The three unknowns ($\eta_L, z_S, H_0$) clearly cannot be solved simultaneously from two constraints, that is, there is a degeneracy among the three unknowns.

Let us illustrate a simple example of the degeneracy among the physical source position $\eta_L$, the source redshift $z_S$ and the Hubble constant $H_0$.
For instance, following the setup of \cite{Huang_2023} and assuming flat $\Lambda$CDM model, with lens redshift and mass being $z_L=0.5,  M_L=1\times 10^{12} M_{\odot}$ respectively, the set of values $\{\eta_L, z_S, H_0\}=\{5 \text{ kpc}, 1, 67.66 \text{ km s}^{-1}\text{ Mpc}^{-1}\}$ lead to $d_\text{Lum}=6776 \text{ Mpc}$ and $y=0.372$. 
However, $\{\eta_L, z_S, H_0\}=\{4.73\text{ kpc},1.5,111.45 \text{ km s}^{-1}\text{ Mpc}^{-1}\}$  also lead to same values of $d_\text{Lum}$ and $y$.
Effectively, it implies the two different sets of $\{\eta_L, z_S, H_0\}$ will also produce the same observed time delay and effective luminosity distance for the lensed GWs.
Therefore, the Bayesian parameter estimation results presented in Fig. 4 of \cite{Huang_2023}, where $z_S, H_0, \eta_L$ are all constrained locally, should not be reliable.
In particular, the Markov Chain Monte Carlo results (in Fig. 4) seem to show an apparent resolution to this degeneracy, which we argued to be incorrect.
Interestingly, their FIM results (indicated by the red lines) appear to be in agreement with our conclusion here, and stands against the work's claim, as it indicates a degeneracy among $\{z_S, H_0, \eta_L\}$ (in Fig. 4).

We note that, this degeneracy is a subset of the similarity transformation degeneracy discussed in the context of gravitational wave lensing (we refer to \cite{poon2024galaxylensreconstructionbased} for the full details and formalism).
The similarity transformation degeneracy states that, irrespective of lens model, in dark siren GW lensing (i.e., no complimentary information from lensing of light), there is a degeneracy between the Einstein radius $\theta_E$, source redshift $z_S$, lens redshift $z_S$ and Hubble constant $H_0$. 
In short, ideally, lens modelling can only resolve the dimensionless lens parameters (such as the Singular Isothermal Ellipsoid axis ratio), the dimensionless source position $\vec{y}$ and the two scaling quantities: the time delay scaling $T_*=\frac{1+z_L}{c}\frac{D_L D_S}{D_{LS}}\theta_E^2$ and the original luminosity distance $d_\text{Lum}$. 
The four parameters $z_L, z_S, \theta_E, H_0$ remain coupled and degenerate inside $T_*(z_L, z_S, \theta_E, H_0)$ and $d_\text{Lum}(z_S, H_0)$.
However, if lens redshift $z_L$ is constrained independently, having additional independent measurement on either source redshift $z_S$ or Einstein radius $\theta_E$ will enable us to measure the Hubble constant $H_0$ \footnote{If the lens model is more complex, or the lens suffers from mass sheet degeneracy, Hubble constant may remain unresolvable.}.

\acknowledgments
Jason S.C. Poon is supported by the Hong Kong PhD Fellowship Scheme (HKPFS) from the Hong Kong Research Grants Council (RGC). 
Jason S.C. Poon and Otto A. Hannuksela acknowledge support by grants from the Research Grants Council of Hong Kong (Project No. CUHK 14304622, 14307923 and  14307724), the start-up grant from the Chinese University of Hong Kong, and the Direct Grant for Research from the Research Committee of The Chinese University of Hong Kong.

\end{document}